\documentclass[10pt]{amsart}
\usepackage{graphicx}

\begin{document}

\title{Genoogle: \\an indexed and parallelized search engine for similar DNA sequences}
\author{Felipe Fernandes Albrecht}
\email[F.~Albrecht]{felipe.albrecht@gmail.com}
\date{\today}

\begin{abstract}
The search for similar genetic sequences is one of the main bioinformatics tasks.
The genetic sequences data banks are growing exponentially and the searching techniques
that use linear time are not capable to do the search in the required time anymore.
Another problem is that the clock speed of the  modern processors are not growing as it did before,
instead, the processing capacity is growing with the addiction of more
processing cores and the techniques which does not use parallel computing does not
have benefits from these extra cores.
This work aims to use data indexing techniques to reduce the searching process computation cost
united with the parallelization of the searching techniques to use the computational capacity
of the multi core processors.
To verify the viability of using these two techniques simultaneously,
a software which uses parallelization techniques with inverted indexes was developed.

Experiments were executed to analyze the performance gain when parallelism is utilized,
the search time gain, and also the quality of the results when it compared with others searching tools.
The results of these experiments were promising, the parallelism gain overcame
the expected speedup, the searching time was $20$ times faster than the parallelized \emph{NCBI BLAST},
and the searching results showed a good quality when compared with this tool.

The software source code is available at https://github.com/felipealbrecht/Genoogle .

\end{abstract}

\maketitle

\section{Introduction}
One of the most important tasks at the bioinformatics is the search for similar genetic
sequences in the data banks. With the new sequencing technologies, the size of these genetics
data banks are growing exponentially~\cite{genbank:07}, and consequently the search time is growing too.

The alignment algorithms, \emph{Needleman-Wunsch}~\cite{needleman:70} and
\emph{Smith-Waterman}~\cite{smith:81}  are algorithms of the dynamic programming
class  \cite{cormen:01}.
They are very sensible alignments algorithms, but their computation and memory costs
are quadratic $O(mn)$~(where $m$ is the input sequence length and $n$ the data bank length).
This computation and memory costs turn them impractical at large data banks sets.
To address this problem, heuristics was developed to reduce the memory and processing consumption of the similar sequences searching processes.
Among the algorithms which use heuristics for similar sequences searching,
the \emph{FASTA} \cite{pearson:88} and \emph{BLAST}\cite{altschul:90,altschul:97,cameron:07} algorithms are the more used and know algorithms.
These algorithms search for areas which has similarities, called \emph{HSP} (High Scoring Pairs), and then, they make the alignment of the best
\emph{HSP} found using a dynamic programming algorithm.
The \emph{FASTA} and \emph{BLAST} algorithms also optimized the alignment cost, because the alignment
is made between words which have a previously detected similarity only.
Nevertheless, the problem complexity continues being $O(nmq)$ ($q$ is the quantity of
sequences in the data banks) because it still necessary to read the data bank sequences
entirely to find the \emph{HSPs}.

The searching process optimization can be made using inverted index to localize \emph{HSPs}.
Inverted indexes are data structures which allow to find the localization of the
indexed data at constant time ($O(1)$).
This type of indexes are utilized at Information Retrieval area, especially in web searching engine,
as \emph{Google}, \emph{Yahoo}, and libraries for data indexing, as the \emph{Apache Lucene} \cite{lucene:09}.

For the search for similar genetic sequences, the sub-sequences of the data bank sequences are indexed,
eliminating the necessity of finding the \emph{HSPs} thought a linear search.
It is quite similar with the indexes found at the end of the books, where rather than to have the text entry,
it have a sub-sequences entries informing where these sub-sequence can be found.
The two most common data structures which are usually used to index the genetic sequences data bank are the
suffixes trees and vectors. Suffixes tress are used at the work of Gusfield~\cite{gusfield:97},
this data structure allows to access the sub-sequence position in linear time,
but the real virtue is to find repeating sequences regions and to obtain the longest common ancestral.
One example of suffixes tress application is presented by \cite{giladi:01},
where an algorithm for building tree to find near-exact sequences is utilized.
At this example, the access time gain has a high memory consumption cost.
At \cite{ning:01} is said that the Delcher~et~al.~\cite{delcher:99} has a memory spent of
$37$~bytes by base at her implementation when using suffixes trees.
As comparison, using the human genome with approximately three billions base pairs,
are necessary more than $103$~gigabytes to store the suffix tree.
At \cite{delcher:02} an optimization is shown in relation to previous work,
where the algorithm is three times faster and it uses one third of the memory.
Even with this reduction, it is necessary more than $34$~gigabytes
to store the suffix tree to index the human genome.

Vector is a data structure which is mainly an array of elements,
where each position of this array represents an information and inside this position,
have another array informing where this information can be found.
Some similar searching sequences techniques that uses vectors as inverted indexes are:
\emph{SSAHA}\cite{ning:01}, \emph{BLAT}\cite{kent:02}, \emph{PatternHunter}\cite{ma:02}
the \emph{miBLAST}\cite{kim:05}, \emph{Megablast}\cite{tan:05}, \emph{MegaBlast}\cite{morgulis:08},
and Kalafus~\cite{kalafus:04} which uses hash tables to align whole genomes.
Transforming based methods are considered out of the scope of this work,
but Jing~\cite{jiang:07} presents methods that use this technique.

Another possible way to reduce the searching time is through parallel computing.
The importance of the parallelization has grown along with the growing multiprocessing capabilities,
like the number of the cores, in the modern processors.
Some tools for similar sequences searching have ways to use these multiprocessing technologies.
As one example, it is possible to inform the \emph{NCBI BLAST} to divide the data bank and
perform parallel searches at these data banks fragments.
As the data bank is divided and the search parallelized, the computational complexity is $O(\frac{nmq}{f})$
($f$ is the quantity of fragments that the data bank will be divided),
or it means, even with the search parallelization, it still have a linear computation complexity.

The computational complexity can be reduced using inverted index,
but searching at the literature, it was not found any technique which uses indexing with parallel programming,
and consequently they do not use the capacity of moderns the multi-core processors.
Thus, the objective of this work is to use the index techniques to optimize the searching
process along with the use of the multi-core processor to reduce the searching time
of the similar genetic sequences.
The developed software, called \emph{Genoogle} uses indexing techniques with inverted index,
index search parallelization, data bank division and parallelization,
bit level sequences codification and alignment algorithms optimizations.
This software was developed utilizing the Java environment and it has a web page, web services
and text mode interfaces.

\section{Methods}
This work presents the \emph{Genoogle}, that is a similar sequence searching engine software
which has as objective to execute fast and with good sensibility searches.
To achieve these goals, it uses inverted index to find nimbly the \emph{HSPs}
and it also uses parallelization techniques to use the multi-core processors capabilities.
\emph{Genoogle} works similarly to the \emph{BLAST},
where given an input sequence, a parameter set, and a data bank where the search will be made,
the software returns the similar sequences from the given input and parameters.

Defining a genetic sequence as a sequence where $\Sigma$~=~\{$A$, $C$, $G$, $T$\}~(\emph{DNA}),
$\Sigma$=\{$A$, $C$, $G$, $U$\}~(\emph{RNA}) and a sub-sequence is a sequence wich is contained
partiality of fully into other sequence.
The sequences at \emph{Genoogle} are divided into fixed length sub-sequences,
where the length is defined by the user before the data bank formatting,
and them, they are codified using $2$ bits for each sequence base.
The sub-sequences are stored as a bit vector into $32$ bits integer and the length can vary from $1$ to $16$ bases.
Changing this value impacts at the search speed and sensibility,
as big is the value, as fast will be the searching process, but the sensibility will be lower.
To save memory, not overlapping windows are utilized to codify the data bank sequences,
but for input sequences, to have more sensibility during the searching process, overlapped windows are utilized.
These codified sequences, with other sequence information: name, identification code,
and description are stored into a file disk, composing the data bank sequences.

Masks are used at the indexed sub-sequences to improve the searching sensibility of the inverted index.
The masks are based on the \emph{PatternHunter}~\cite{ma:02} work.
The masks inform which sub-sequences bases should be maintained or removed and, consequently
they increase the probability to find sub-sequences at the index.
The masks provide two gains:
the sensibility, which allows to search at the index with not-exact sequences,
and it saves index space, because longer sub-sequences will be transformed into smaller,
having less index entries and less sub-sequences at the data bank.

\emph{Genoogle} has some run time parameters which can improve the sensibility or the
search performance. The parameters for the sensibility are:
the maximum distance between index entries for be considered for the same \emph{HSP},
the minimum \emph{HSP} size and the drop off for sequences extension. Changing these parameters
impact on the sensibility, allows to find more \emph{HSP}, but it is expected to have
more false positives and to slow down the performance.

\subsection{Inverted Indexes}
\emph{Genoogle} use vectors as inverted indexes data structures.
The size of the main vector is the quantity of possible sub-sequences:
as the \emph{DNA} alphabet has $4$ letters, it is possible $4^n$ sub-sequences, being $n$ the
defined sub-sequences length, having the inverted index $4^n$ entries.
The size of each sub-sequence vector varies according the quantity of this sub-sequence into the data bank.
Each sub-sequence occurrence uses two integers of $4$ bytes to store the information:
One integer is used to store the identifier of the sequence and
the other integer is used to store the position at the sequence.
The inverted index uses $4$ bytes to the sequences identifier and more $4$ to
the position at this sequence, turning possible to index approximately
$4,25$~billions~($2^{32}$) sequences and each one can have until this length,
being the available memory the major limit to the quantity of sequences and their size.

\begin{figure} [ht]
\centering
 \fbox{
 \includegraphics[width=0.60\textwidth]{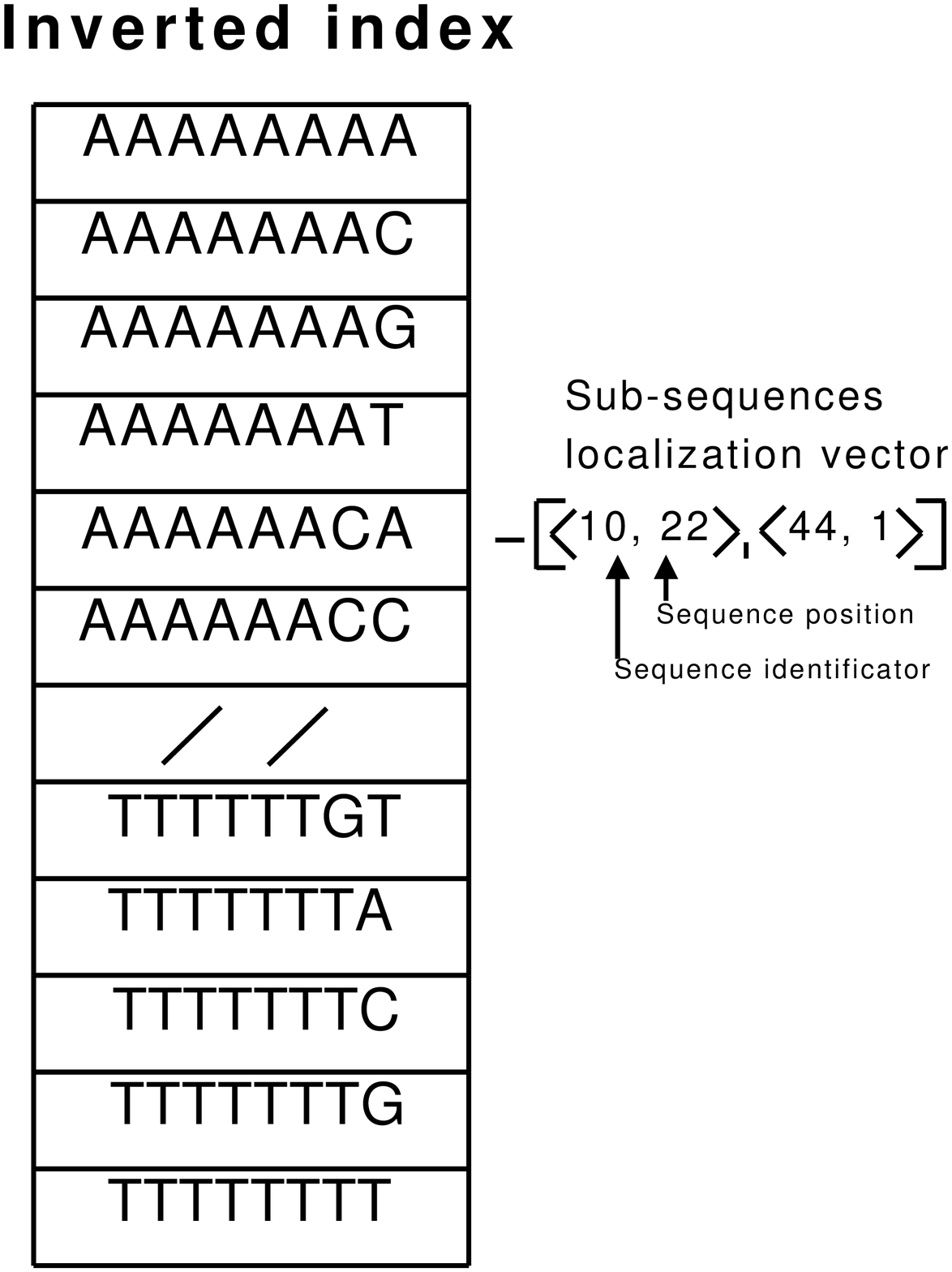}
 }
 \caption{Sub-sequences inverted index structure.}
 \label{fig:indice}
\end{figure}

The mask are applied at the data bank sequences during the indexing process.
They are read and divided into sub-sequences,
at each one the mask is applied on its sub-sequences and hence these masked sub-sequences are indexed.
Using masks allow having longer sub-sequences with smaller sensibility loose.
For example, using the mask \emph{111010010100110111}, where \emph{1} means that the
base in that location should be preserved and the \emph{0} means that the base there
should be removed, the sub-sequences read from the data bank will have $18$ bases,
and after the mask be applied, will have $11$ bases.
In this way, it is used a vector to index $11$ bases sub-sequences, but
the sequences are divided into $18$ bases sub-sequences,
saving inverted index structure memory.

\emph{Genoogle} needs approximately $(( (l/m) s) 8) + ((4 ^ s) 16)$ bytes to store the inverted index.
Being $l$ the length of the data bank in bases, $m$, the total mask length, and $s$ the sub-sequences
length.
For a data bank with $4$ billions bases and sub-sequences with $11$ bases, there will be $363$ of millions sub-sequences,
requiring $2.833$~megabytes to store the inverted index.
Using masks with $18$ bases length, there will be $222$ millions sub-sequences and the inverted
index will need $1.759$~megabytes, resulting on a gain of $30\%$ of the total memory required.
To build the inverted indexes it is used the sort-based method~\cite{witten:99} and the inverted index
and the formatted data bank are stored at the disk. During the \emph{Genoogle} start time,
the whole inverted index is read and loaded into the main memory,
the data bank meta informations, like file offset, are also loaded into the main memory,
but the data bank sequences are read from the disk when the sequences informations are necessary.

\subsection{Searching process}
After that the inverted index and the data bank meta informations are loaded into the main memory,
\emph{Genoogle} is ready to run the searches. The searching process is divided into $7$ phases:
\begin{itemize}
 \item Input sequence processing;
 \item Index searching for similar sub-sequences and construction of the \emph{HSPs};
 \item \emph{HSPs} extension and merge;
 \item Merging overlapped \emph{HSPs};
 \item Selection of the high scored \emph{HSPs};
 \item Local alignment of the selected \emph{HSPs};
 \item Selection and exhibition of the best alignments.
\end{itemize}

\begin{figure} [ht]
\centering
\fbox{
\includegraphics[width=0.60\textwidth]{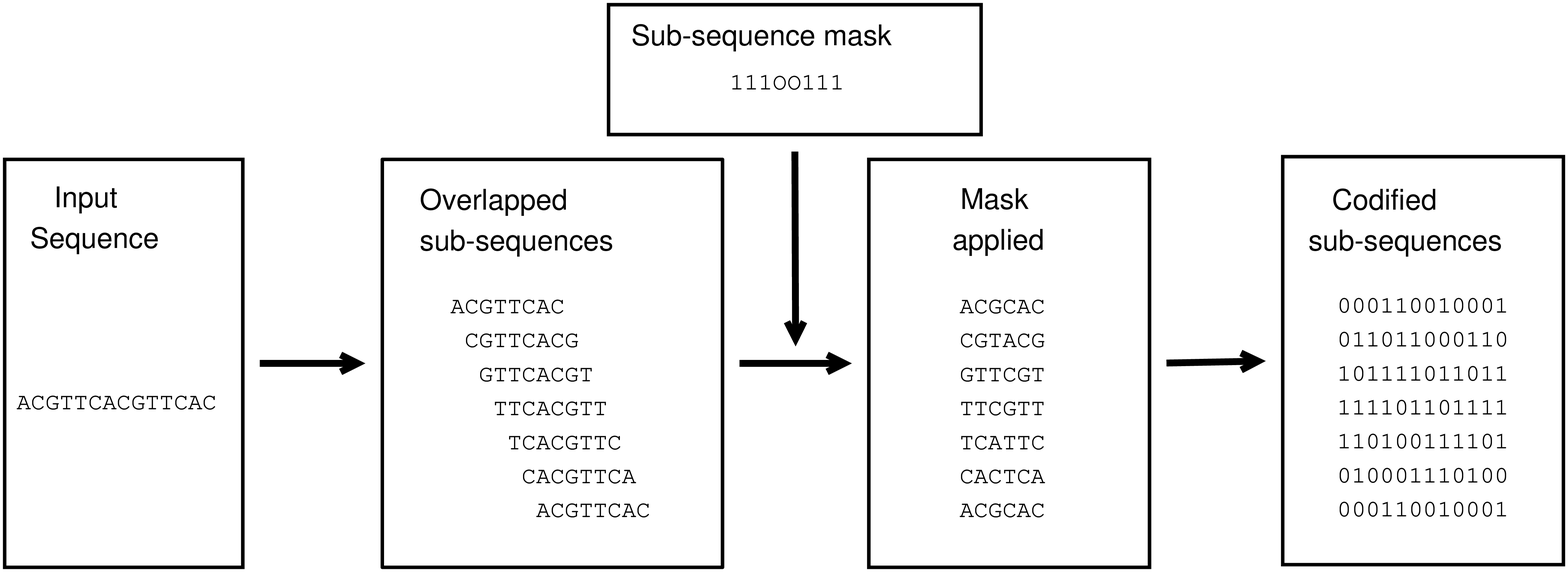}
}
\caption{Input sequence processing.}
\label{fig:input_sequence_processing}
\end{figure}

The input sequence processing firstly applies the mask at each overlapped sub-sequence of the input sequence
and codify the resulting sub-sequence to the binary representation used by \emph{Genoogle}.
The input sequence processing is shown at Fig.~\ref{fig:input_sequence_processing}.
As the input sub-sequences are codified as binary data into an integer,
it is possible to obtain the sub-sequence value directly from this data.
Because the determined sub-sequence position at the index is its own value
it turns the index searching process simpler and direct.

\begin{figure} [ht]
\centering
\fbox{
\includegraphics[width=0.60\textwidth]{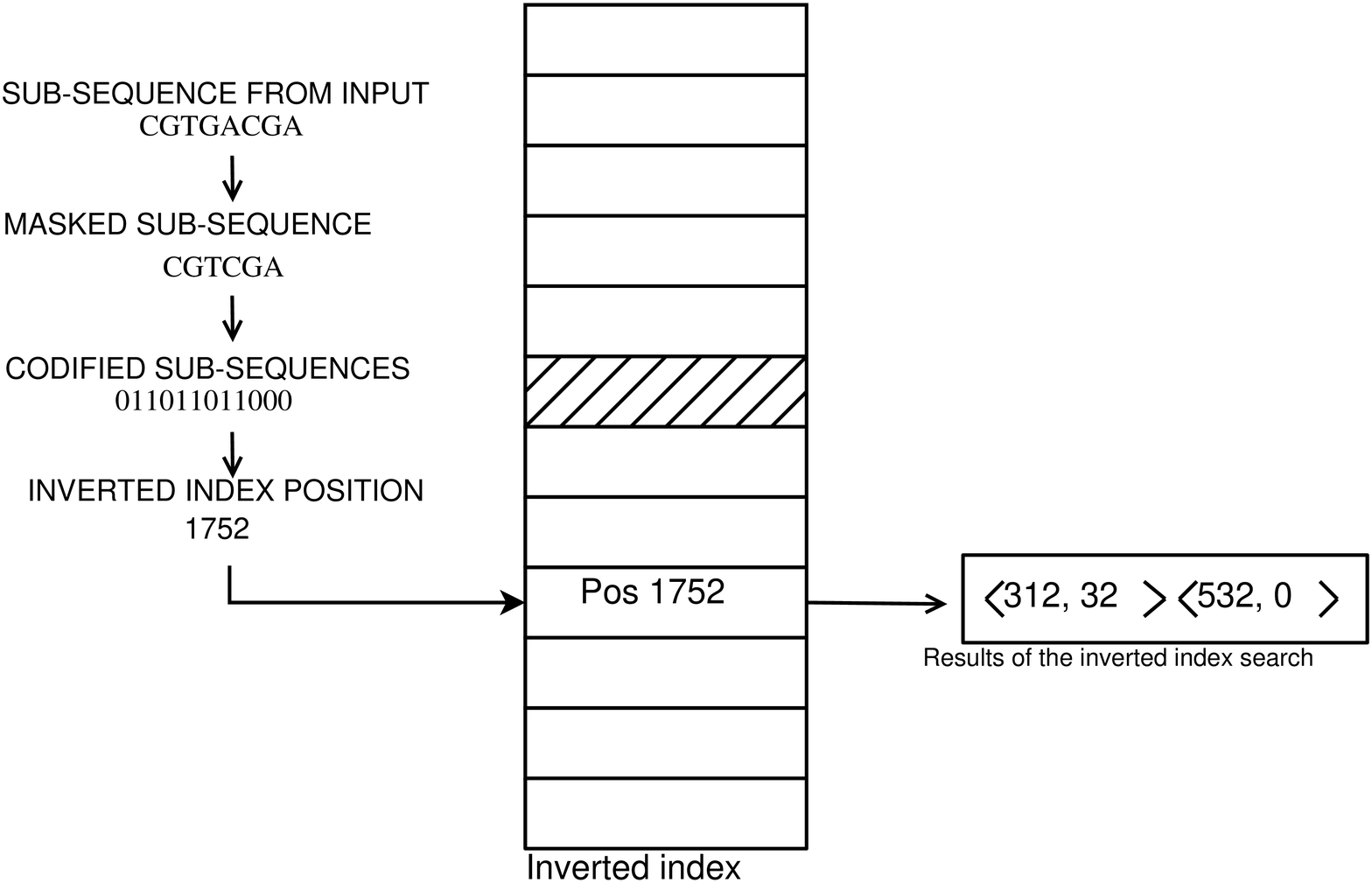}
}
\caption{Obtaining sub-sequence localization data from the inverted index.}
\label{fig:sub_obtain}
\end{figure}

At Fig.~\ref{fig:sub_obtain} is shown the process to retrieve the informations from the inverted index.
For each masked and encoded sub-sequence from the input sequence, using its encoded value, is retrieved from
the inverted index all places which have this sub-sequence at the data bank sequences.
The retrieved informations are stored into an array of arrays, where each position represents a data bank sequence.
If two or more retrieved information are are closer than a specified parameter, they are merged into one retrieved information.
These informations are filtered by their length and the remaining ones are them called High Scoring Pairs (\emph{HSP}).
The \emph{HSP} have five information:
Initial and final positions at the input sequence and at the data bank sequence, and
the length of this area, where it gets the length of the \emph{HSP} in relation of
the data bank and of the input sequence and get the smaller value.

After the index searching phase, the located \emph{HSPs} are extended to the both directions to try to enlarge their length.
During the extension phase, two or more \emph{HSPs} which were closed can be overlapped, generating duplicate results.
Hence, it is verified after the extension phase if the \emph{HSPs} have overlapped regions,
and if they do, they are merged into one new \emph{HSP}.

After the extension and merging phase, the \emph{HSPs} are sorted by their lengths in a decreasing way.
It is possible to specify a parameter to inform the maximum quantity of alignments which should be returned.
The objective of this parameter is to reduce the number of alignments built,
saving processing and returning only the most significant alignments,
it is also possible to set to return all \emph{HSPs} found.

After the selection, it is made a local alignment for the selected \emph{HSPs}.
The local alignment is built by a modified version of the \emph{Smith-Waterman} algorithm.
The alignment algorithm is modified in order to limit the distance from the main diagonal
which will be used to produce the alignment.
Rather than to use the whole matrix to build the alignment,
it is used only the closest cells of the main diagonal.
This limit is used because the two sub-sequences which will be aligned have a high similarity,
turning unnecessary to calculate the whole alignment matrix.
To reduce the memory use, was also used a technique to divide the alignment matrix in smaller matrices.
This technique divides the two sequences into segments, and each segment is aligned separately.
After the segments alignment, the results are merged and considered the final result of whole sequences.
After the \emph{HSPs} alignments, they are sorted by the alignment score, and the results
are returned with other information, as the e-value, score, normalized score , and alignment
position in the input and data bank sequence, to the user.

\section{Parallelization Methods}

To improve the searching time and to use the multi-processing capabilities,
\emph{Genoogle} uses three parallelization techniques:
inverted index access parallelization, extension and alignment parallelization,
and data bank division parallelization.

The data bank division parallelization is made dividing the data bank in fragments, similarly how \emph{NCBI BLAST} does.
The indexes searches and \emph{HSPs} construction are made independently into threads for each data bank fragment.
After locating the \emph{HSPs} by the individual threads, they are extended, merged, sorted, filtered,
aligned, sorted by their score, and returned to the user.
This technique is interesting for large data banks,
but this parallelization technique does not parallelize the whole searching process:
the extension, sorting and alignment phases are not parallelized by this technique.

Analyzing the searching data bank fragmentation process,
it was verified that not every searching thread have the same execution time.
It happens because some fragments have more similar sequences than others.
Analyzing the searching time, it was verified that the index searching time
represents about $60\%$ of the whole searching time.
About $30\%$ of the searching time is for the \emph{HSPs} alignments,
because it, a component was developed to parallelize the sequences
extension, sorting and alignment using all available computer processing core.

At this parallelization technique, the threads search the input sub-sequences
at the data bank fragments index and then put the \emph{HSPs} into a
collection shared by all index searching threads.
After the index searching phase, the \emph{HSPs} collection is sorted by the \emph{HSP} length and
the longest are selected using a parameter that informs how many
alignments should be returned.
The remaining \emph{HSPs} are put into a \emph{FIFO} queue, where it has
extenders and aligners which will perform the extension and alignment.
These extenders and aligners use independent threads, they read a \emph{HSP} from the queue,
them perform the extension and alignment and put the result into another shared collection.
These threads are managed by an executor, when a thread finish its job,
the manager sends to it another \emph{HSP} to extend and align until all \emph{HSPs}
are extended and aligned. It is possible to configure the quantity of simultaneous threads.

The memory required by the inverted index structure is the main problem
of the data bank division.
For example: using $11$ base pairs length sub-sequences,
they are necessary $32$ megabytes to store only the inverted index structure.
Theoretically, to use the whole computational capacity of a
computer with $8$ processing cores, it is necessary to divide the
data bank in $8$ parts, becoming necessary to use up $512$ megabytes to store only
the inverted index structure.
Because of the memory requirement for the data bank division, it is important to use
a complementary approach.
Along with the data bank division, the input sequences are also
divided and performed the search parallel. Thus it is possible to parallelize the
searching process without overloading the memory with more data structures.

\begin{figure*}[ht]
\centering
\fbox{
\includegraphics[width=0.95\textwidth]{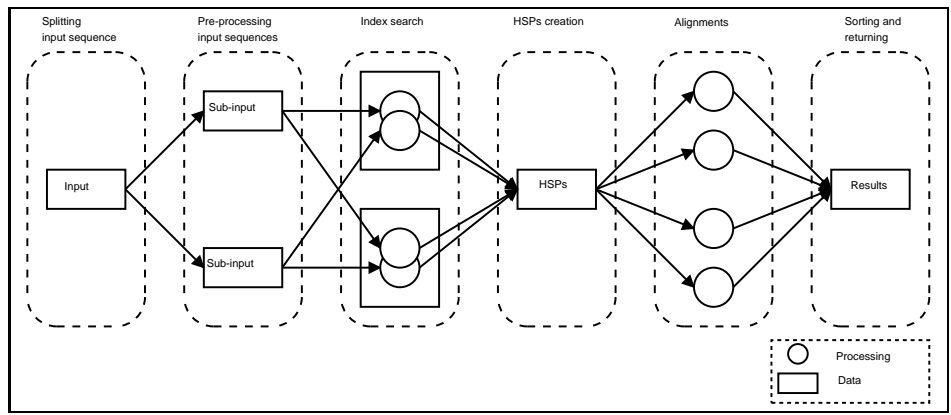}
}
\label{fig:parallel_search}
\caption{Obtaining sub-sequence localization data from the inverted index.}
\end{figure*}

This parallelization divides the input sequence in sub-inputs,
and it searches each sub-input at the inverted index independently.
It is also useful because it also parallelize the input query processing.
After the index search, the \emph{HSPs} that are from the same data bank sequence
and are closer, are merged into one \emph{HSP}.
Using the data bank division in two fragments and the input sequence in two sub-inputs,
they are used $4$ threads to search at the inverted index, being the memory overload
of only two inverted indexes structures.
At the Fig~\ref{fig:parallel_search} is shown the complete parallelized \emph{Genoogle} search
dividing the input query in two sub-inputs and the data bank in two fragments.

\section{Implementation}

\emph{Genoogle} is developed using the Java Environment version $1.6$.
The Java Environment was choose because it is multi-platform and it has a framework and primitives for parallel computing.
The library \emph{BioJava}~\cite{holland:08} was used during the first part of the development, but now it was removed from the project.
At the first implementations, \emph{BioJava} was used to reading, parsing, storing and genetic sequences alignment,
but it was verified that the reading and parsing methods require too much memory, so,
new and optimized classes was developed to perform these tasks.

The main user interface is an text mode interface, where the user types the search command and the results are
stored into a \emph{XML} file that is defined by the user.
This interface has commands to perform the search, to list the available data banks, to obtain the parameters
list, to run the garbage collection, to run the last command executed, and to run a batch file containing commands.
The batch command is interesting because with it is possible to write a file with all commands that should
be executed and to inform the \emph{Genoogle} to execute them without any used intervention during the execution
of these commands.

\emph{Genoogle} has also an embedded simple web page interface, best suitable for testing,
which contains only one input field, for the input sequence and a button to perform the search.
The search results of the web page interface are a \emph{XML} document formatted and shown as a \emph{HTML} web page by a \emph{XSL} document.
Together with the web page, \emph{Genoogle} has a web-services interface, implemented using \emph{JAX-WS}.
Using the web-services interface, it is possible to execute queries, to set parameters, retrieve the data bank available list,
and others tasks inside of a programming script. The users can write scripts to access \emph{Genoogle} services automatically
using their preferred programming language and perform their searches without manual intervention.

\section{Results}
For the experiments was utilized a data bank with sequences of the phase $3$ of the human genome project~\cite{NCBI:08}
along with RefSeq~\cite{RefSeq:08} data banks.
The RefSeq data banks were used because they are verified and they have high quality.
The human genome project data bank is the \emph{hs\_phase3} and has approximately $3.8$~$Gb$.
The RefSeq data bank are the \emph{cow} with approximately $57$~$Mb$, \emph{frog} with $20$~$Mb$,
\emph{human} with $112$~$Mb$, \emph{mouse} with $93$~$Mb$, \emph{rat} with $73$~$Mb$, and
\emph{zebrafish} with $62$~$Mb$, totalizing approximately $417$~$Mb$.
At the experiments was used the data banks from the RefSeq along with the \emph{hs\_phase3},
totalizing $4.25$~$Gb$ and being necessary approximately $2$ gigabytes of memory to
store it.

For the execution of the experiments, they were generated $11$ sets of input sequences.
Each set has $11$ sequences with approximately the same size, being $1$ sequence got from the
data bank and $10$ are mutations of these sequences.
The sets are with sequences with $80$ base pair (bp), $200bp$, $500bp$, $1.000bp$,
$5.000bp$, $10.000bp$, $50.000bp$, $100.000bp$, $500.000bp$ e $1.000.000bp$.
The searches were made using the data bank parallelism, dividing the input sequence
parallelism and extending and aligning the sequences simultaneously.

At the experiments it was used a computer with $16$ gigabytes of RAM,
Linux version $2.6.18$ and
 Java Environment version $1.6$ with the \emph{JVM}~\emph{JRockit}~version $3.0.3$.

\begin{figure} [ht]
\centering
\fbox{
 \includegraphics[width=0.30\textwidth,angle=270]{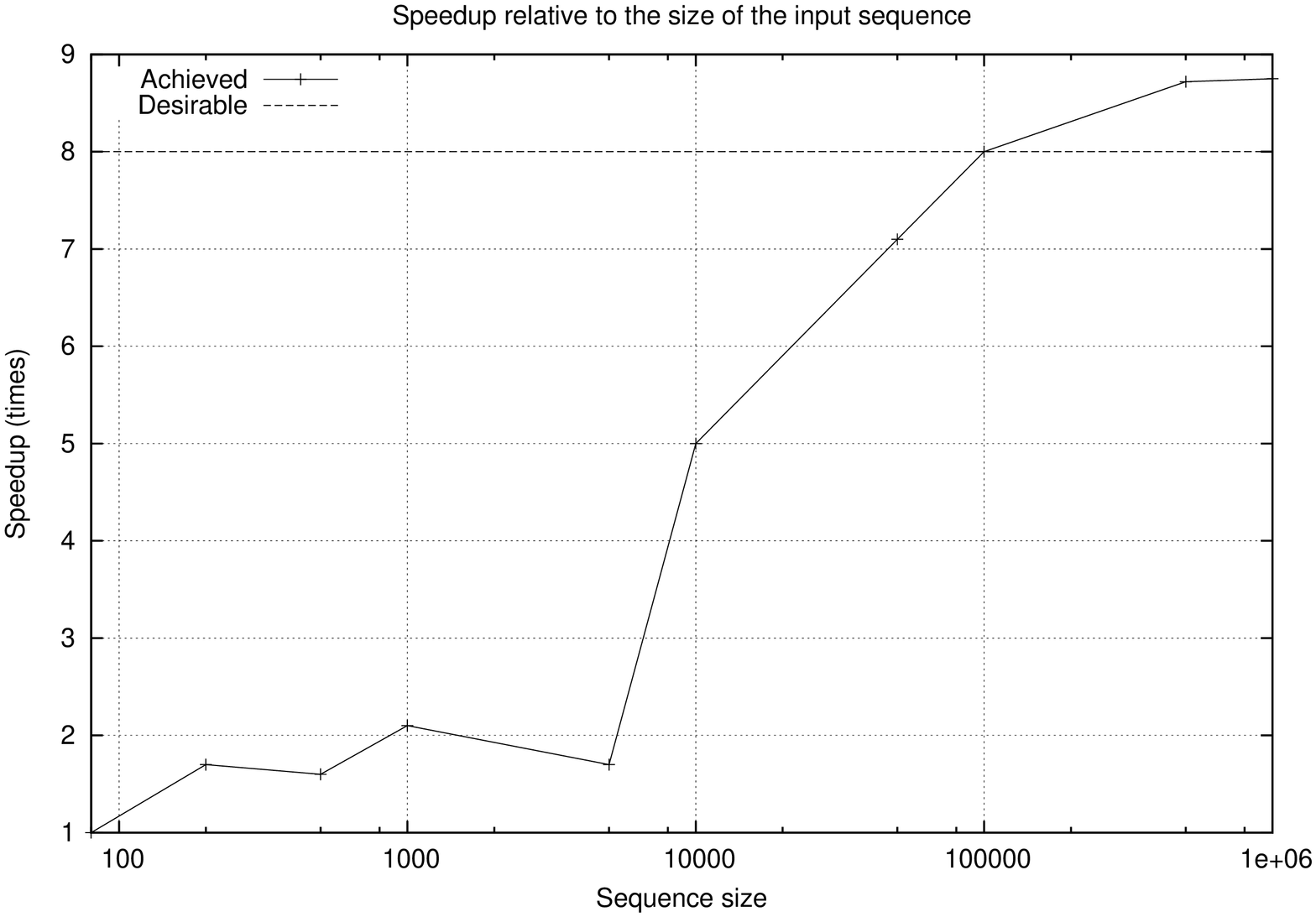}
} \caption{Input sequence speedup relation.}
\label{fig:speedup}
\end{figure}

For entries up to $5.000bp$ the gain using the parallelism is low, giving a gain of only $2$ times
and the total time increases when the input sequence is divided into more than $4$ parts.
It happens because the searching time for these small input sequences are too low, less than $200ms$,
and the synchronization overload impacts directly to the searching time.
For input sequences with $10.000bp$ it there is a gain of $5$ times with the utilization of the parallelization techniques.
For inputs with $50.000bp$ and up, the speedup are $8$, that is the aiming gain, and with $500.000bp$ and $1.000.000bp$
inputs the gains overcome this speedup.

The use of tools that use suffix trees was discarded for the time comparison because of the memory required by these tools.
Thought tests, it was verified that the \emph{BLAT}~software can not handle data banks bigger than $4$~$Gb$.
\emph{MegaBlast} was not verified because \cite{ma:02} says that \emph{MegaBLAST} was developed to be efficient
on the searching time, but the results quality is worst than \emph{NCBI BLAST},
it happens because the seeds minimum size is $28bp$.
Because of the lower results quality, it was decided not to execute experiments with this tool.
The \emph{Indexed MegaBlast}~\cite{morgulis:08} can not be executed because its memory requiring $4$ times the data bank, needing more than the $16$~$Gb$ available.
And finally, it was not possible to obtain the \emph{PatternHunter} to execute the experiments.
Thus, it was decided to compare the performance only against the \emph{NCBI BLAST}.

Comparing the sequential search times of \emph{Genoogle} and \emph{BLAST} it is shown that Genoogle is
almost $20$ times faster and comparing the parallel times, Genoogle is $26,60$ times faster.
It is interesting to realize that for smaller input, until $5.000bp$, the time gains in relation with \emph{BLAST} are not so good
because the parallelization techniques does not use all their potential in these small inputs.
The time difference at bigger inputs comes until $29$ times for $100.000bp$ input.
It is important to realize that the sequential version of Genoogle is faster than the parallel executions of the \emph{BLAST}.

\begin{small}
\begin{table}[ht]
\begin{center}
\begin{tabular}{|r|r|r|r|}
\hline
Base pairs &		BLAST (ms) & 	Genoogle (ms)	& 	Gain (times)	 \\
$80$	   	& 		$5.572$	   & 	$150$			&	$37,00$	\\
$200$ 	   	&		$8.882$   & 	$460$			&	$19,30$	\\
$500$ 	   	&		$14.488$  & 	$340$			&	$42,61$	\\
$1.000$	   	&		$19.087$  & 	$570$			&	$33,48$	\\
$5.000$	   	&		$58.902$  & 	$2.400$			&	$24,54$	\\
$10.000$   	&		$98.160$  & 	$5.318$			&	$18,45$	\\
$50.000$   	&		$604.785$  &	$31.499$		& 	$19,20$	\\
$100.000$  	&		$1.973.333$  &	$75.610$		&	$26,09$	\\
$500.000$ 	&		$7.700.571$   &	$393.450$		&	$19,57$	\\
$1.000.000$	&		$1.229.988$   &	$76.909$		&	$16,00$	\\
\hline
Total 		& 		$11.713.768$  &   $586.706$     &   $19,96$   \\
\hline
\end{tabular}
\caption{Time comparison between sequential NCBI BLAST and sequential Genoogle.}
\label{tab:comp_blast_genoogle_1}
\end{center}
\end{table}
\end{small}

\begin{small}
\begin{table}[ht]
\begin{center}
\begin{tabular}{|r|r|r|r|}
\hline
Base pairs  & BLAST (ms)   & Genoogle (ms)	&   Gain (times) \\
$80$ 	    &     $1.061$  &	$150$  		&	$7,00$	\\
$200$  		&     $2.145$  &	$270$		&	$7,94$	\\
$500$  		&     $3.170$  &	$210$		&	$15,09$	\\
$1.000$ 	&     $2.853$  &	$270$		&	$10,56$	\\
$5.000$ 	&    $10.387$  &	$1.341$		&	$7,74$	\\
$10.000$   	&    $13.027$  &	$1.050$		&	$12,40$	\\
$50.000$   	&    $78.067$  &	$4.440$   	&	$17,58$	\\
$100.000$  	&   $276.779$  &	$9.380$		&	$29,50$	\\
$500.000$ 	& $1.206.212$  &	$45.120$ 	&	$26,73$	\\
$1.000.000$ &   $193.090$  & 	$8.780$ 	&	$22,00$ \\
\hline
Total  &  $1.786.791$  		& 	 $67.011$ 	& 	$26,66$	\\
\hline
\end{tabular}
\caption{Time comparison between parallel NCBI BLAST and parallel Genoogle.}
\label{tab:comp_blast_genoogle_8}
\end{center}
\end{table}
\end{small}


\subsection{Results quality}

The results quality was analyzed comparing the Genoogle results against the BLAST results and verifying
which \emph{HSPs} were identified as similar and what is the percentage of \emph{HSPs} that were identified by \emph{BLAST}
and not by the Genoogle.
For each input sequences, a collection with the found alignments by BLAST was created
and it is verified if these alignments were found by Genoogle. It is accounted how many alignments
were found and it is generated a percentage for each \emph{E-Value} range
varying from $10e^{-90}$ a $10e^0$.

Following is shown the graphic which shows the proportion of alignments found by Genoogle in relation to the BLAST
according the alignment \emph{E-Value}.
This graphic was generated from the data where it was verified which of the alignments found by the BLAST were found by the Genoogle too.
In this graphic it is possible to observe that until the \emph{E-value} $10e^{-35}$ more than $90\%$ of the alignments
were found by the \emph{Genoogle}.
Until the \emph{E-Value} $10e^-{15}$ more than $60\%$ of the alignments were found and with the \emph{E-Value}
$10e^{-10}$ almost $55\%$ of the alignments were found.
Above this \emph{E-Value} the quantity of alignments found is bellow $40\%$.

\begin{figure} [ht]
\centering
\fbox{
 \includegraphics[width=0.30\textwidth,angle=270]{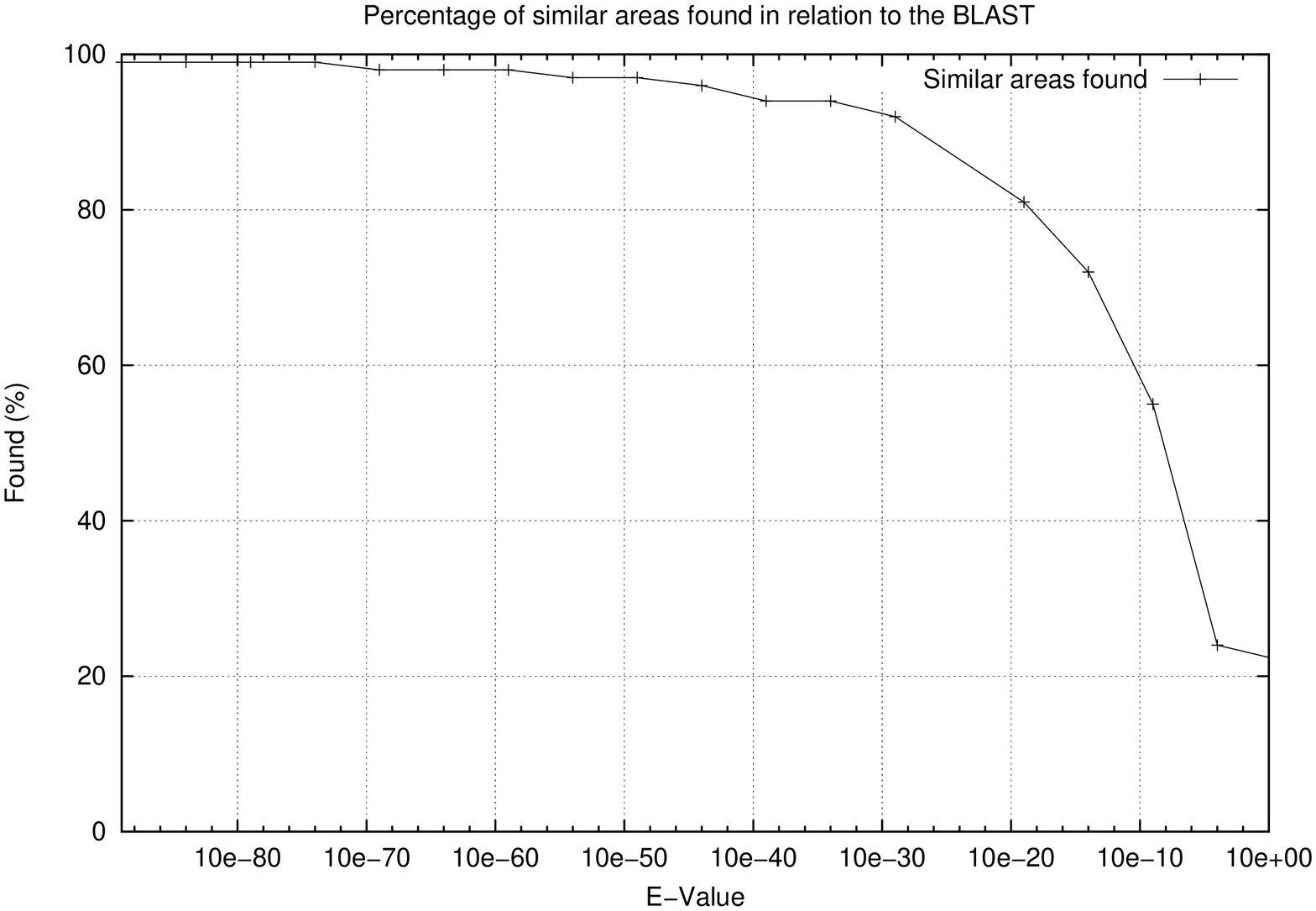}
} \caption{Proportion of alignments found by Genoogle in relation to BLAST.}
\label{fig:found_evalue}
\end{figure}

Analyzing this graphic is realized that the Genoogle is a good tool to find alignments
with the \emph{E-Value} lower than $10e^{-25}$.
It happens because alignments with this \emph{E-Value} are usually long, with more than $100bp$,
and consequently are found easily.
From the \emph{E-Value} $10e^{-30}$ there is a drop in the results quality, where it is stabilized close
to the \emph{E-Value} $1$.
Alignments with the \emph{E-Value} highest than $0.005$ are alignments which can not be possible to
infer a close homology~\cite{brass:98}.
This, it can be observed that Genoogle has a very good search quality until the maximum $10e^{-20}$ \emph{E-Value},
but it has a quality drop until $10e^{-5}$.
The alignments should not to be considered to homology inference for values higher than $10e^{-4}$.

The experiments showed that the Genoogle quality results is comparable to the BLAST when the alignment
\emph{E-Value} is representative, demonstrating a possible homology between the two aligned sequences.
More sensible search can be archived changing the searching parameters,
as the maximum distance between the sub-sequences information got from the index, and the minimum
HSP length.
Changing the minimum HSP length for alignments with \emph{E-Value} higher than $1e^{-5}$,
the HSP found for this \emph{E-Value} grew to approximately $80\%$ and the search
time was raised only $3\%$.

\section{Conclusion}
This work presented a genetic similarity sequences searching software which uses
data bank sequences indexing along with parallel computing.
To ensure the effectiveness and the search quality of to use index and parallel computing was
developed and implemented the Genoogle tool.
This software was implemented using the Java $1.6$ and it can be executed at Windows, Linux,
and Mac environment.
Experiments were executed to verify the results execution time and the results quality.
The searching time was really good, with speedup of more than $20$ times in relation to parallelized BLAST.
The results quality was good, finding relevant alignments, but it can be optimized by changing the
searching parameters.
Thus, merging the indexing techniques with the three parallelization techniques and the option to
optimize the search configurations, Genoogle proved to be an effective tool and its results
have good quality.

As the main contributions of this work, it should be noted primarily as the first tool
in the literature to do the genetic sequences search using indexing and parallel computing.
Considering that parallel computing importance has increased with increasing number of cores
at the processors and also the importance of the data indexing to optimize the searching process
in data banks wich has an exponential grow, this work has a relevance for addressing these two
issues together.

The software is available at \emph{genoogle.pih.bio.br} and there is a demonstration page at \emph{pih.bio.br:8080}.


\end{document}